\documentclass[aps,preprintnumbers,superscriptaddress,twocolumn]{revtex4}

\usepackage{amsfonts,amssymb,amsmath}            
\usepackage{graphics}            
\usepackage{graphicx}
\usepackage{pstricks}
\usepackage{amsthm}                              
\newcommand{\XX}{$XY$\ }  
\def\id{\mathbb{I}}
\newcommand{\ket}[1]{\left | #1 \right\rangle}
\newcommand{\bra}[1]{\left \langle #1 \right |}
\newcommand{\proj}[1]{\ket{#1}\bra{#1}}

\newcommand{\half}{\mbox{$\textstyle \frac{1}{2}$}}
\newcommand{\h}[1]{\mathcal{H}_{#1} }

\newcommand{\be}{\begin{equation}}
\newcommand{\ee}{\end{equation}}
\newcommand{\bea}{\begin{eqnarray}}
\newcommand{\eea}{\end{eqnarray}}
\newcommand{\expec}{\mathbb{E}}

\theoremstyle{plain}

\theoremstyle{definition}

\newcommand{\beas}{\begin{eqnarray*}}
\newcommand{\eeas}{\end{eqnarray*}}
\newcommand{\Tr}{\text{Tr}}
\def\id{\leavevmode\hbox{\small1\kern-3.8pt\normalsize1}}

\newcommand{\ux}{\underline{x}}

\newcommand{\ue}{\underline{e}}

\def\reff#1{(\ref{#1})}

\begin{document}

\title{Perfect Transfer of Arbitrary States in Quantum Spin Networks}

\date{\today}

\author{Matthias \surname{Christandl}}
\email[]{matthias.christandl@qubit.org} \affiliation{Centre for Quantum
Computation,
             DAMTP,
             Centre for Mathematical Sciences,
             University of Cambridge,
             Wilberforce Road,
             Cambridge CB3 0WA, UK}

\author{Nilanjana \surname{Datta}}
\email[]{n.datta@statslab.cam.ac.uk} \affiliation{Statistical Laboratory,
             Centre for Mathematical Science,
             University of Cambridge,
             Wilberforce Road,
             Cambridge CB3 0WB, UK }
\author{Tony C. \surname{Dorlas}}
\email[]{dorlas@stp.dias.ie}
\affiliation{Dublin Institute for Advanced Studies,
School of Theoretical Physics,
10 Burlington Road, Dublin 4, Ireland.
}
\author{Artur \surname{Ekert}}
\email[]{artur.ekert@qubit.org} \affiliation{Centre for Quantum Computation,
             DAMTP,
             Centre for Mathematical Sciences,
             University of Cambridge,
             Wilberforce Road,
             Cambridge CB3 0WA, UK}
\affiliation{Department of Physics,
             National University of Singapore,
             Singapore 117\,542, Singapore}

\author{Alastair \surname{Kay}}
\email[]{alastair.kay@qubit.org} \affiliation{Centre for Quantum Computation,
             DAMTP,
             Centre for Mathematical Sciences,
             University of Cambridge,
             Wilberforce Road,
             Cambridge CB3 0WA, UK}

\author{Andrew J. \surname{Landahl}}
\email[]{alandahl@mit.edu} \affiliation{Center for Bits and Atoms,
             Massachusetts Institute of Technology,
             Cambridge, MA 02139, USA}
\affiliation{HP Labs,
             Palo Alto, CA 94304-1126, USA}

\begin{abstract}

We propose a class of qubit networks that admit perfect state
transfer of any two-dimensional quantum state in a fixed period of
time. We further show that such networks can distribute arbitrary
entangled states between two distant parties, and can, by using
such systems in parallel, transmit the higher dimensional systems
states across the network. Unlike many other schemes for quantum
computation and communication, these networks do not require qubit
couplings to be switched on and off.  When restricted to $N$-qubit
spin networks of identical qubit couplings, we show that $2\log_3
N$ is the maximal perfect communication distance for hypercube
geometries. Moreover, if one allows fixed but different couplings
between the qubits then perfect state transfer can be achieved
over arbitrarily long distances in a linear chain. This paper
expands and extends the work done in \cite{Christandl}.
\end{abstract}

\maketitle

\section{Introduction}

An important task in quantum information processing is the
transfer of
quantum states from one location ($A$) to another location ($B$).
In a quantum communication scenario this is rather explicit, since the
goal is
the communication between distant parties $A$ and $B$ (e.g., by means of
photon
transmission). Equally, in the interior of quantum computers good
communication
between different parts of the system is essential. The need is thus to
transfer quantum states and generate entanglement between different
regions contained within the system.
There are various physical systems which can serve
as quantum channels, one of them being a quantum spin system. This
 can be generally defined as a collection of interacting
qubits on a graph, whose dynamics is
governed by a suitable Hamiltonian, e.g., the Heisenberg or
$XY$--Hamiltonian.
One way to accomplish this task is by
multiple
applications of controlled swap operations along the communication line.
Every
external manipulation, however, inevitably induces noise in the system.
It is
therefore desirable to minimise the amount of external control on the
system, to the point that they do not require any external manipulation
whatsoever.

Quantum communication over short distances through a spin chain,
in which adjacent qubits are coupled by equal strength has been
studied in detail, and an expression for the fidelity of quantum
state transfer has been obtained \cite{Bos03, Sub03}. Similarly, in \cite{Shi}, near perfect state transfer was achieved for uniform couplings provided a spatially varying magnetic field was introduced. The
propagation of quantum information in rings has also been investigated \cite{OL04}.

In our work we focus on the situation in which state transfer is
{\em{perfect}}, i.e., the fidelity is unity, and in which we can
design networks such that this can be achieved over arbitrarily
long distances. We will also consider the case in which \emph{no}
external control is required during the state transfer, i.e., we
consider the case in which we have, after manufacturing the
network, no further control over its dynamics. In general this
will lead us to think about more complicated networks than
the linear chain or chains with pre-engineered nearest--neighbour
interaction strengths. We provide two alternative methods for understanding how perfect state transfer is achieved with pre-engineered couplings.
 This paper expands and extends the work
done in \cite{Christandl}. Subsequent work has examined the extension of this problem to higher excitation subspaces \cite{Christandl:2004a}. The subject of perfect state transfer has been independently studied in the first and second excitation subspaces in \cite{Lambropoulos}, where its implementation in an array of quantum dots was considered.

More specifically, we address the problem of arranging $N$
interacting qubits in a network such that it allows for perfect
transfer of any quantum state over the longest possible distance.
Two qubits are coupled via an $XY$-interaction if an edge connects
the two corresponding sites. We show further how one can use these
networks to transfer entangled quantum states and to generate
entanglement between distant sites in the network. The connection
between our approach and the continuous-time quantum random walk
is highlighted and, in a particular example, contrasted with the
corresponding result for the classical continuous time random
walk.

The paper is organised as follows: In Section
\ref{Section-State-Transfer} we set the scene by introducing the
problem of perfect state transfer in quantum spin systems. In
Section \ref{Section-Conditions} we derive a necessary condition
for our problem in the case of graphs with mirror symmetry.
This is used in Section \ref{Section-Limitations} to give a limitation of the transfer in chains with uniform couplings.
In Section \ref{Section-Greater} we
investigate hypercube geometry as a way to enlarge the previously
found limit and compare our result to a classical analog in
Section \ref{Section-Classical}. The quantum walk on the hypercube
is found useful to derive a \emph{modulated} spin chain in
Section \ref{Section-Projection} that allows perfect transfer over
arbitrary distances. We will exhibit a group-theoretic
interpretation of this chain in Section \ref{Section-Arbitrary}.
In Sections \ref{Section-Entanglement-Transfer} and
\ref{Section-Phase} we consider applications for entanglement
transfer and the introduction of arbitrary phase gates
\emph{on-the-fly}.

\section{State Transfer in Quantum Spin Systems}

\label{Section-State-Transfer} In order to set the scene, let us
first consider quantum state transfer over a general quantum
network. We define a general finite quantum network to be a
simple, connected, finite graph $G:=\{V(G),E(G)\}$, where $V(G)$
denotes the finite set of its vertices and $E(G)$ the set of its
edges. Two vertices $i, j \in V(G)$ are adjacent if $(i,j) \in
E(G)$. To any such graph $G$ one can associate an adjacency matrix
$A(G)$ whose elements satisfy
\be
A_{ij}(G):=\begin{cases} 1\hbox{ if } (i,j) \in E(G)\\
0\hbox{ otherwise.}\end{cases}
\ee
A quantum spin system associated with such a graph is defined by attaching a spin--$\half$ particle to each vertex of the graph. To each vertex $i \in V(G)$ we can
therefore associate a Hilbert space $\h{i} \simeq \mathbb{C}^2$. The Hilbert
space associated with $G$ is then given by
\be \h{G}=\bigotimes_{i \in V(G)} \h{i}=\left( \mathbb{C}^2 \right)^{\otimes
N}, \label{hilbert} \ee
where $N:=|V(G)|$ denotes the total number of vertices in $G$.

We define the distance, $d(i,j)$, between any two vertices $i, j \in V(G)$ to
be the number of edges of the shortest path between $i$
and $j$, i.e., the graph geodesic between the two vertices.

Consider the dynamics of the system to be governed by the
quantum--mechanical Hamiltonian \be H_G=\half\sum_{(i,j) \in E(G)}
J_{ij}\Bigl[ \sigma_i^x \sigma_j^x+\sigma_i^y \sigma_j^y \Bigr].
\label{ham} \ee We use the symbols $\sigma_i^x, \sigma_i^y$ and
$\sigma_i^z $ to denote the familiar Pauli matrices acting on the
on--site Hilbert space ${\cal{H}}_i$ and $J_{ij}$ is the coupling
strength between the $i^{th}$ and $j^{th}$ sites on the graph.
Note that $J_{ij}=J_{ji}$ since $H_G$ is Hermitian. The total
$z$-component of the spin, given by \be \sigma_{tot}^z:= \sum_{i
\in V(G)} \sigma_i^z \label{spintot} \ee is conserved, i.e., $
[\sigma_{tot}^z, H_G]=0.$ Hence the Hilbert space $\h{G}$
decomposes into invariant subspaces, each of which is a distinct
eigenspace of the operator $\sigma_{tot}^z$.

For the purpose of quantum state transfer, it suffices to restrict
our attention to the $N$--dimensional eigenspace of
$\sigma_{tot}^z$, which corresponds to the eigenvalue $(2-N)/2$.
Let us denote this subspace by ${\cal S}_G$. Initial quantum
states that are in this subspace will remain there under time
evolution. A basis state in ${\cal S}_G$ corresponds to a spin
configuration in which all the spins except one are down and one
spin is up. Such a basis state can hence be denoted by the ket
$|j\rangle$, where $j$ is the vertex in $G$ at which the spin is
up. Thus $\{|j\rangle \,| \,j \in V(G)\}$ denotes a complete set
of orthonormal basis vectors spanning ${\cal S}_G$.

When restricted to the subspace ${\cal S}_G$, $H_G$ is
represented, in the above--mentioned basis, by an $N \times N$
matrix which is {\em{identical}} to the adjacency matrix $A(G)$ of
the underlying graph $G$ \footnote{In \cite{Osb03} it was shown,
starting from $G$, how to construct graphs whose adjacency matrix
governs the time evolution in higher excitation subspaces}. The
time evolution of the system under the action of the Hamiltonian
$H_G$ can be interpreted as a \emph{continuous time quantum walk}
on $G$ (first considered by Farhi and Gutmann in 1998 \cite{FG98};
see also \cite{ChildsGutmann:2002}). This is because the latter is
defined as the time--evolution in an $N$--dimensional Hilbert
space spanned by states $\{ |j \rangle\}$, where $j \in V(G)$,
with a Hamiltonian given by the adjacency matrix of $G$.

The spin system on the graph $G$ described above plays the role of a
(noiseless) quantum channel. We see below that the continuous time
quantum walk on $G$ can be viewed as a quantum state transfer along the
channel.

The process of transmitting a quantum state from $A$ to $B$ proceeds in four
steps:
\begin{enumerate}
\item Initialisation of the spin system to the state
$|{{\underline{0}}}\rangle:=|0_A0 \cdots 00_B\rangle$, which
corresponds to the configuration of all spins down. This state is
a zero energy eigenstate of $H_G$ \footnote{Experimentally, this
initialization can, for example, be achieved by application of a magnetic
field to align the spins.}.
\item Creation of the quantum state $|\psi\rangle_A \in \h{A}$ (at
vertex $A$) which is to be transmitted. Let $|\psi\rangle_A =
\alpha |0\rangle_A + \beta |1\rangle_A$ with $\alpha,\beta \in
\mathbb{C}$ and $|\alpha|^2+|\beta|^2=1. $
\item Time evolution
of the system for an interval of time, say $t_0$.
\item Recovery of the state at vertex $B$, the latter being given by the reduced density matrix
$\rho_B$ acting on ${\cal{H}}_B$.%
\end{enumerate}
The state of the entire spin system after step $2$ is given by
\bea
|\Psi(t=0)\rangle &=& |\psi_A 00 \cdots 0 0_B \rangle \nonumber \\
&=&  \alpha|0_A00 \cdots 00_B \rangle +\beta |1_A00 \cdots 00_B \rangle\nonumber\\
& =& \alpha |{\underline{0}} \rangle + \beta |1\rangle,
\label{input}
\eea%
It evolves in time $t$ to

\be |\Psi(t)\rangle= \alpha |{{\underline{0}}}\rangle+\sum_{j=1}^N
\beta_j(t) |j\rangle
\label{evolution} \ee%
with complex coefficients $\alpha, \beta_j(t)$, where
$1=|\alpha|^2+\sum_{j=1}^N|\beta_j(t)|^2$. The initial conditions
are given by $\beta_A(0) = \beta$ and $\beta_j(0) = 0$ for all
$j\ne A$. The coefficient $\alpha$ does not change in time, as
$\ket{{\underline{0}}}$ is the zero-energy eigenstate of $H_G$.
Hence, it does not even acquire a phase factor during the
evolution of the state.

The output state at $B$ after a time $t$ is given by the reduced density matrix
\bea
\rho_B(t):&=&\Tr_{\h{G \setminus \{B\}}} |\Psi(t)\rangle \langle\Psi(t)| \nonumber\\
&=& \left(%
\begin{array}{cc}
  1 - |\beta_B(t)|^2 &  \alpha\beta^*_B(t) \\
  \alpha^*\beta_B(t)& |\beta_B(t)|^2\\
\end{array}%
\right). \eea A measure of the overlap between the input state,
$\rho_A := \proj{\psi}$, and the output state is the fidelity,
\begin{equation}
\begin{split}
F(\rho_A,\rho_B(t))
    :&= \Tr\sqrt{\rho_A^{1/2}\rho_B(t)^{\phantom{1}}\rho_A^{1/2}}\\
    &= \sqrt{\bra{\psi}\rho_B(t)\ket{\psi}} \\
    &=\sqrt{|\alpha|^2 \left(1\! -\!2|\beta_B|^2\!+\!\beta_B\beta^*\!
+ \!\beta_B^*\beta\right)\!+\!|\beta_B|^2}
\end{split}
\label{fid}
\end{equation}
where it is understood that $\beta_B$ depends on $t$.

Since the $\ket{0}_A$ component of the state $\ket{\psi}_A$ is
invariant under the evolution, it suffices to focus on the
evolution of the $\ket{1}_A$ component of the state i.e., to the
choice $\alpha =0$ and $\beta = 1$ in \reff{input}. It is
therefore convenient to consider the \emph{transfer fidelity} \be
f_{AB}(t):=\beta_B(t) \equiv \langle B| e^{-iH_Gt} |A\rangle.
\label{fid2} \ee where $\ket{A}\equiv \ket{1}=\ket{1_A0 \ldots
00_B}$ and $\ket{B}\equiv \ket{N}=\ket{0_A0 \ldots 01_B}$ and we
have taken $\hbar=1$.

Here we focus only on {\it{perfect state transfer}}. This means
that we consider the condition
\be |f_{AB}(t_0)|=1 \quad {\hbox{
for some}}\quad 0< t_0 <\infty \label{eq:mod_fidel} \ee
which we
interpret to be the signature of perfect communication (or perfect
state transfer) between $A$ and $B$ in time $t_0$. The effect of
the modulus in \reff{eq:mod_fidel} is that the state at $B$, after
transmission, will no longer be $\ket{\psi}$, but will be of the
form \be \alpha\ket{0}+e^{i\phi}\beta\ket{1}. \ee The phase factor
$e^{i\phi}$ is not a problem because $\phi$ is independent of
$\alpha$ and $\beta$ and will thus be a known quantity for the
graph, which we can correct for with an appropriate phase gate.

The perfect communication distance $d(A,B)$ is given by the
distance on the graph, for which perfect state transfer is
possible. For a fixed number of qubits, $N$, our aim is to find
quantum networks which maximise $d(A,B)$. We achieve this in two
different ways:
\begin{enumerate}
\item By fixing the nearest--neighbour couplings to be identical but considering more complicated graphs (see Sections \ref{Section-Limitations} and \ref{Section-Greater}).
\item By considering linear chains but allowing the nearest--neighbour couplings to be different (Sections \ref{Section-Projection} and \ref{Section-Arbitrary}).
\end{enumerate}

Note that if there is perfect communication between $A$ and $B$ in
a time $t_0$, then perfect communication also occurs for all times
$t$ satisfying
\be t=(2n+1)t_0,  \,\hbox{ where }\, n \in
\mathbb{Z}, \ee provided the graph is mirror--symmetric (see
Section \ref{Section-Conditions}).

\section{Conditions for Perfect State Transfer in Systems with Mirror Symmetry}
\label{Section-Conditions}

In the rest of the paper, we will examine different graphs for the
purposes of perfect state transfer. These graphs will have mirror
symmetry. By mirror symmetry, we mean that the graph is identical
from the points of view of $A$ and $B$. So, a linear chain with $A$ and
$B$ at opposite ends is an example of such a system. The obvious
question is how can we tell if a proposed graph will permit
perfect state transfer? A necessary condition, as we will show, is
that the ratios of differences of the eigenvalues of the
Hamiltonian, $H_G$, must be rational provided the graph is mirror
symmetric.

With a system capable of perfect state transfer, initialised in the state $\ket{A}$, at time $t_0$ we have the state
\be
e^{-iH_Gt_0}\ket{A}=e^{i\phi}\ket{B}
\ee
but by the definition of a symmetric system, $A$ and $B$ are entirely equivalent, and thus after another period of time $t_0$, we have the state
\be
e^{-i H_G 2 t_0} \ket{A}=e^{-iH_Gt_0}e^{i\phi}\ket{B}=e^{i2\phi}\ket{A}
\ee
and thus the system is periodic, up to a phase $2\phi$, with period $2t_0$. Thus we conclude that {\em{a mirror symmetric system must be periodic if it is to allow perfect state transfer}}. This may be written most simply as
\be
|\bra{A}e^{-iH_G2t_0}\ket{A}|=1
\ee
for some time $0<t_0<\infty$.

Let us examine the general state of a periodic system with period
$2t_0$. We can write
\be
\ket{\psi(2t_0)}=\sum_j a_j e^{-i2E_j t_0}\ket{j}=e^{i2\phi}\sum_j
a_j\ket{j},
\ee
for eigenstates $\ket{j}$ of $H_G$ with corresponding eigenvalues
$E_j$. Hence for all of the stationary states $\ket{i}$, we have
the condition
\be
2E_it_0-2\phi=2k_i\pi
\ee
where the $k_i$'s are integers. Eliminating $\phi$ between two of these, we get that
\be
(E_i-E_j)2t_0=2\pi(k_i-k_j)
\ee
and eliminating the $t_0$ between any two of these ($E_{i'}\neq E_{j'}$) gives
\be
\frac{E_i-E_j}{E_{i'}-E_{j'}}=\frac{k_i-k_j}{k_{i'}-k_{j'}}\in \mathbb{Q},
\label{must_be_rational}
\ee
where $\mathbb{Q}$ denotes the set of rational numbers. As the $k_i$'s are integers, this implies that the ratio is rational.
Hence, a symmetric system capable of perfect state transfer must be periodic, which is equivalent to the requirement that the ratios of the differences of the eigenvalues are rational.

\section{Limitations for Perfect Communication of a Uniformly
Coupled Chain} \label{Section-Limitations}

It is desirable to maximise the distance over which communication is possible for a fixed number of qubits. The optimal arrangement, in this case, is just a linear chain of $N$ qubits, where $A$ and $B$ are the qubits at opposite ends of the chain.

Let us start with the \XX chain of qubits, with uniform couplings $J_{i,i+1}=1$ for all $1\leq i\leq N-1$. The Hamiltonian reads
\be
H=\half\sum_{i=1}^{N-1}\sigma_i^x\sigma_{i+1}^x+\sigma_i^y\sigma_{i+1}^y .
\ee
In this case, one can
compute $f_{AB}(t)$ explicitly by diagonalizing the Hamiltonian or the
corresponding adjacency matrix. The eigenstates and the corresponding
eigenvalues are given by
\bea
\tilde{\ket{k}} &=& \sqrt{\frac{2}{N+1}} \sum_{n=1}^N
\label{eq:states}
\sin\left(\frac{\pi k n}{N+1}\right)\,\ket{n} \\
E_k & = & - 2\, \cos \frac{k \pi}{N+1},
\label{eq:eigstates}%
\eea%
with $k=1, \ldots, N$.  Thus
\be%
f_{AB}(t)=\frac{2}{N+1}\sum_{k=1}^N \sin\Big(\frac{\pi k }{N+1}\Big) \sin
\Big(\frac{\pi k N}{N+1}\Big)e^{-i E_k t}.
\label{eq:two_link}
\ee%

Perfect state transfer from one end of the chain to another is possible for $N=2$ and $N=3$, where we find that $f_{AB}(t)=-i\sin(t)$ and
$f_{AB}(t)=-\left[\sin\left(\frac{t}{\sqrt{2}}\right) \right]^2$ respectively.

We have shown that perfect state transfer is possible for chains containing 2 or 3 qubits. We will now prove that it is not possible to get perfect state transfer for $N\geq 4$.

A chain is symmetric about its centre. Hence the condition Eq. \reff{must_be_rational} for perfect state transfer applies, i.e.
\be
\frac{E_m-E_n}{E_{m'}-E_{n'}}
\in \mathbb{Q}
\ee
where the $E_m$'s are eigenvalues of the unmodulated chain, as given in \reff{eq:eigstates}. We will explicitly show that there is a set of eigenvalues for which this expression does not hold for all $N\geq 4$.

We are free to choose any values for the indices (provided $E_{m'}\neq E_{n'}$), so let us choose that $m=2$, $n=N-1$, $m'=1$ and $n'=N$. Hence we see, using \reff{eq:eigstates}, that we require
\be
\frac{\cos{\frac{2\pi}{N+1}}}{\cos{\frac{\pi}{N+1}}} \in \mathbb{Q}
\ee
to hold for perfect state transfer.
To find the values of $N$ for which this holds, we make use of algebraic numbers.
An \emph{algebraic number} $x$ is a complex number that satisfies an equation of the form
\be \label{poly} a_0 x^n +a_1 x^{n-1} + \cdots a_{n-1}x + a_n=0, \ee
with integral coefficients $a_i$. Every algebraic number $\alpha$ satisfies a
unique polynomial equation of least degree. The degree of this
polynomial is called the \emph{degree} of $\alpha$.

If $\alpha$ satisfies a monic polynomial (i.e., a polynomial with $a_0=1$) then it is
called an \emph{algebraic integer} of degree $n$. Note that an algebraic integer of degree $n$ is also an algebraic number of degree $n$.
Rational numbers are algebraic numbers with degree $1$, and numbers with degree
$\geq 2$ are irrational.

Lehmer proved (for example, see \cite{irrational}) that if $N>1$ and $\gcd(k,N+1)=1$, then $\cos \left(\pi k/(N+1)\right)$ is an algebraic integer of
degree $\phi(2(N+1))/2$, where $\phi$ is the Euler phi--function.

For $n\geq 3$ it can be shown \cite{EulerPhiBound} that
\be \phi(n)\geq \frac{n}{e^\gamma \log \log n+\frac{3}{\log \log n}}
\label{eqn:star}
\ee
holds, with $\gamma \approx 0.5772$, Euler's constant.
Using this bound, and by inspection of values not covered by the bound, we see that
$\phi(2(N+1))/2 \geq 3 \text{ for } N \geq 6$.

We need to prove that if $\cos \theta$ is an algebraic number of degree $\geq 3$, the
quotient
\[ \frac{\cos 2 \theta}{\cos  \theta} \] is irrational, where $\theta=\pi/(N+1)$.

Assume that this expression is rational, i.e.
\be
\frac{\cos 2 \theta}{\cos  \theta} =\frac{p}{q} \quad p,q\in\mathbb{Z}.
\ee
Using the trigonometric identity
\be \cos 2\theta =2 \cos^2 \theta-1\ee
we can write
\be (\cos \theta )^2-\frac{p}{2q}\cos \theta -\half=0 \ee
which has rational coefficients. According to the definition, $\cos \theta$
is therefore algebraic with degree $\leq 2$. Given that, from \reff{eqn:star}, $\cos\theta$ is an algebraic number of degree $\geq 3$, then we have a contradiction and therefore $(\cos 2\theta)/\cos \theta$ must be irrational.

Hence we see that for $N \geq 6$ perfect state transfer is impossible because $\text{deg}(N)\geq 3$. This simply leaves $N=4$ and $N=5$ unproved, which can be done by straightforward evaluation.
Thus for $N \geq 4$
\[ \frac{\cos 2 \theta}{\cos \theta}\notin \mathbb{Q}. \]
Hence, perfect state transfer is impossible for unmodulated chains of length $N\geq 4$.

\section{Perfect State Transfer Over Greater Distances}
\label{Section-Greater}

Perfect state transfer over arbitrary distances is
impossible for a simple unmodulated spin chain. Clearly it is
desirable to find a graph that allows state transfer over larger
distances, and to that end we examine the $d$--fold Cartesian
product of the two--link (three--vertex) chain, $G$. We denote this by $G^d$.

In general the Cartesian product of two graphs $G:=\{V(G), E(G)\}$
and $H:=\{V(H), E(H)\}$ is a graph $G \times H$ whose vertex set
is $V(G) \times V(H)$ and two of its vertices $(g,h)$ and
$(g',h')$ are adjacent if and only if one of the following hold:

\noindent (i) $g=g'$ and $\{h,h'\} \in E(H)$

\noindent (ii) $h=h'$ and $\{g,g'\} \in E(G)$.

Let $A = (1,1,1,\ldots,1)$ and $B =(3,3,3,\ldots,3)$ denote the antipodal
points of $G^d$.

We prove that for {\em{any dimension}} $d$
\be|f_{AB}(t)| =1, \quad {\hbox{for}} \quad t = t_0 =  \frac{\pi}{\sqrt{2}}.\ee
Hence, $t_0$ is the time for perfect communication between the vertices $A$ and $B$
of $G^d$.

Let $\{\lambda_i(G), 1 \le i \le
|V(G)|\}$ and $\{\lambda_j(H), 1 \le j \le |V(H)|\}$ denote the set of
eigenvalues of the graphs $G$ and $H$
respectively. The eigenvalues of the adjacency matrix of their Cartesian
product $G \times H$ are precisely the numbers: $ \lambda_i(G) +
\lambda_j(H),$ with $1 \le i \le |V(G)|$ and $1 \le j \le
|V(H)|$, where each number is obtained as many times as its multiplicity as
an eigenvalue of the adjacency matrix $A(G\times H)$. This is because
\be
A(G\times H) = A(G) \otimes \id_{V(H)}  + \id_{V(G)} \otimes A(H), \label{tp} \ee
where $\id_{V(H)}$ is the $|V(H)| \times |V(H)|$ identity matrix (see e.g., \cite{graphtheory}).

The eigenvalues of the adjacency matrix of $G^d$ are given by
\be
\{j \sqrt{2}\, | \,j \in \{ 0, \pm 1, \pm 2, \ldots, \pm d \} \} \ee
and therefore the ratios of differences of the eigenvalues are all rational.

As already observed, the Hamiltonian of a system coupled via nearest neighbour \XX interactions is identical to the adjacency matrix. This will hold equally for the Cartesian product of $G$. Hence,
\bea
H=A(G^d)&=&\sum_{j=0}^{d-1}\id^{\otimes j}\otimes A(G)\otimes\id^{\otimes d-j-1}  \\
e^{-iHt}&=&\left(e^{-iA(G)t}\right)^{\otimes d}.
\label{eq:gd}
\eea
Thus, if we select a time $t=\pi/\sqrt{2}$, then we get perfect
state transfer along each dimension. Each term in the tensor
product of \reff{eq:gd} applies to a different element of the
basis (for example, each acts on a different 1 in the definition
of $A$, or a different 3 in $B$). We therefore achieve perfect
state transfer between $A$ and $B$ (as well as between any qubit
and its mirror, such as $(1,1,1,2,3)\rightarrow(3,3,3,2,1)$). The
fidelity of the state transfer is simply the $d^{th}$ power of the
fidelity for the original chain \reff{eq:two_link}.

Perfect transfer of a single qubit state can also be achieved
between the antipodes of a one--link hypercube in any arbitrary
dimension, $d$, in a constant time $t_0=\pi/2$. This is because
perfect transfer occurs across a chain of two qubits in this time.

We can also extend this to the one--link hypercube which is coupled via the Heisenberg interaction. This is because, in the case of a two--qubit chain, the Hamiltonian in the single--excitation subspace is represented by a matrix with identical diagonal elements, and hence is the same as the Hamiltonian of an $XY$ model up to a constant energy shift, which just adds a global phase factor.

Note that
perfect transfer is possible across a ring of 4 spin-1/2
particles. The topology of this is exactly the same as a 2-fold
Cartesian product of a one--link chain, hence it is
a special case of the hypercube we have been discussing (whether it is coupled with the Heisenberg or \XX coupling).

\section{Classical Continuous Time Random Walk on the Hypercube}
\label{Section-Classical}

In the previous Section we showed that for hypercubes generated from both the one--link (two--vertex) chain and the two--link (three--vertex) chain,
the perfect state transfer time, $t_0$, is independent of the dimension $d$.
In this Section we will investigate the behaviour of the mean hitting time of the classical continuous--time symmetric random walk on $G^d$, which we denote by $t_{cl}$, and compare it to $t_0$.
We will focus our attention on the two--link hypercube, since in the quantum case it provides us with a greater communication distance, $d(A,B)$ than the one--link hypercube does. Unlike $t_0$, we show that $t_{cl}$ grows exponentially with the dimension $d$ (equation \reff{result}). We also note that the case of the one--link hypercube has previously been studied \cite{one_link}.

A two--link hypercube $G^d$ is generated by taking the
$d$--fold Cartesian product of the graph $G:=\{V(G), E(G)\}$ where
$V(G)=\{1,2,3\}$ and $E(G)=\{\{1,2\}, \{2,3\}\}$.
Hence, the state space for the classical continuous--time random walk on $G^d$ is $\{1,2,3\}^d$. Transitions are allowed
from a vertex  $\ux \in G^d$ to $\ux\pm {\ue}_i$ where ${\ue}_i$ is
the $i$-th unit vector
and $\ux$ is a $d$--dimensional vector with components $x_i$ i.e., transitions are allowed to all the nearest-neighbours with equal probability.

If $T$ is the random variable
defined as the hitting time of $B$, for a random walk starting at
$A$, then\be
t_{cl}:= \expec (T)
\ee
where $\expec(T)$ denotes the expectation value of $T$.
The random variable $T$ can be written as
\be
T = \sum_{i=1}^N X_i,
\ee
where $N$ is a random variable which gives the number of jumps
that the random walker undergoes in going from $A$ to $B$, and the
$X_i$'s are the holding times between successive jumps. We have
\bea
\expec(T) &=& \expec(\sum_{i=1}^N X_i)\nonumber\\
&=& \expec(N) \expec(X_1)\nonumber\\
&=& \expec(N),
\eea
where we have made use of the fact that the $X_i$'s are independent
and identically distributed with mean $\expec(X_i)= 1$. Note that
$\expec(N)$ is the mean hitting time of the corresponding jump chain, which is
a discrete-time Markov chain. Hence,
to estimate the mean hitting time $t_{cl}$ of the continuous--time
random walk, it suffices to consider the
discrete--time random walk given by the jump chain of the
original walk.

All the information that we need is contained within the
$3^d\times 3^d$ transition matrix, $P$. An element $P_{A,B}$ is
the probability of transition from $A$ to $B$, hence after $N$
steps the probability of hitting $B$ (irrespective of whether it
has previously hit) is $(\underbrace{P\times P \times \ldots
\times P}_{N})_{1,3^d}$, which we shall denote as $P_{1,3^d}^N$,
and hence
\be
\expec(N)=\sum_{n=2d}^\infty
nP_{1,3^d}^n\prod_{m=2d-1}^{n-1}(1-P_{1,3^d}^m).
\label{expectation}
\ee
Since we have a two--link hypercube, it will always take at least $2d$ steps to get from one corner to the opposite one. In fact, because we have a two--link hypercube, it will always take an even number of steps to hit the opposite corner. Thus, $P_{1,3^d}^m=0$ for odd $m$ and $m<2d$.

So, all we are interested in is the element $P_{1,3^d}^{2n}$,
which we expect to tend towards a constant value for large $n$
i.e., $(P^2P^{2n})_{1,3^d}\approx P_{1,3^d}^{2n}$ as $n\rightarrow
\infty$. Thus, we are confronted with the problem of finding the
first element in the eigenvector of $P^2$ which has eigenvalue 1
(see, for example, \cite{markov_chains}).

Take, as an example, the special case of $d=2$ where the first row of
$P^{2n}$ is given by
$$
\left(\begin{array}{ccccccccc}
a_n & 0 & \frac{1}{6} & 0 & \frac{1}{3} & 0 & \frac{1}{6} & 0 & b_n
\end{array}\right)
$$
The 0's occur because in an even number of steps, you cannot get to a point that is odd. For $n=1$, $a_n=\frac{1}{3}$ (the probability of returning to the start node) and $b_n=0$ (the probability of getting to the exit node). The sum of all the elements in the row must be 1, and hence $a_n+b_n=\frac{1}{3}$ for all $n$. We find a recursion relation for $b_n$, $\frac{1}{9}+\frac{1}{3}b_n=b_{n+1}$, and as $n\rightarrow\infty$, we find that $b_n\rightarrow\frac{1}{6}$.

Let us return to the general case. The location of a given node of
the hypercube can be represented by $\ket{x_1,x_2\ldots x_d}$
where $x_i\in\{1,2,3\}$ i.e., $x_i$ specifies, for the $i^{th}$
dimension, which of the three nodes we are positioned on. All of
the properties of a given node depend only on $r$ and $d$, where
$r$ is a count of the number of $x_i=2$. For example, the
transition rate from one node at $r$ to an adjacent node is
\be
P_{\text{nearest neighbour}}=\frac{1}{d+r}
\ee
This quantity can be understood because the transition probability is the same for all connected nodes, and all $x_i=1$ or $x_i=3$ are only connected to $x_i=2$ in the $i^{th}$ dimension, whereas $x_i=2$ has two links.

Given that all properties of a node only depend on $r$ and $d$, that must also be true for the eigenvector, $a$. Hence, we can denote the elements of $a$ by $a_r$. The element of the eigenvector that we are interested in corresponds to the position $\ket{1,1\ldots 1}$, i.e., $r=0$, so we want to find the element $a_0$. We will now find the elements of $a$, the eigenvector of $P^2$.

In two steps, there are only five ways to get to a specific lattice point at a distance $r=2k$.
\begin{enumerate}
\item{Start at that point, make one step away, and then make the same step in reverse. This happens with a probability
\be
p_{\text{return}}=\frac{1}{d+2k}\left(\frac{d-2k}{d+2k+1}+\frac{4k}{d+2k-1}\right).
\ee
}
\item{Start at a distance $2k-2$. There are $4{{2k} \choose 2}$ equivalent jumps, which each occur with probability
\be
\frac{2}{(d+2k-2)(d+2k-1)}.
\ee
}
\item{Start at a distance $2k+2$. There are ${{d-2k} \choose 2}$ of these jumps, each occuring with probability
\be
\frac{2}{(d+2k+2)(d+2k+1)}.
\ee
}
\item{Start at a distance $2k$ and go around two edges of a square (e.g., $\ket{1,2}\rightarrow\ket{1,1}\rightarrow\ket{2,1}$). There are $4k(d-2k)$ points from which this type of move will get us to the specific node we are interested in, and this transition happens with a probability
\be
\frac{1}{d+2k}\left(\frac{1}{d+2k+1}+\frac{1}{d+2k-1}\right).
\ee}
\item{Start at a distance $2k$ and travel the length of a chain (e.g., $\ket{1,2}\rightarrow\ket{2,2}\rightarrow\ket{3,2}$). There are $d-2k$ such jumps, each occuring with probability
\be
\frac{1}{(d+2k)(d+2k+1)}.
\ee}
\end{enumerate}

Knowing these, it is possible to write down the elements $P^2a$, so we can solve for the eigenvector.
\bea
&4{{2k} \choose 2}\frac{2}{(d+2k-2)(d+2k-1)}a_{2k-2}&+  \nonumber\\
&{{d-2k} \choose 2}\frac{2}{(d+2k+2)(d+2k+1)}a_{2k+2}&+ \\
&\frac{1}{d+2k}\left(\frac{(4k+2)(d-2k)}{d+2k+1}+\frac{4k(d-2k+1)}{d+2k-1}\right)a_{2k}&=a_{2k}.    \nonumber
\eea

Starting with the special case of $k=0$, we see that
\be
a_0=\frac{d}{d+2}a_2.
\ee
and the general solution
\be
a_{2k}=\frac{d+2k}{d+2k+2}a_{2k+2}
\ee
can then be proved by induction. We thus have the required elements, and just need to normalise them, remembering that there are $2^{d-2k}{d \choose {2k}}$ identical elements $a_{2k}$.
\be
\sum_{k=0}^{[d/2]}2^{d-2k}{d \choose {2k}}a_{2k}=1
\ee
This gives that
\be
a_{2k}=\frac{d+2k}{2d3^{d-1}}
\ee
and, in particular
\be
a_0=\frac{3}{2}3^{-d}.
\ee
We know that, as $n\rightarrow\infty$, $P_{1,3^d}^{2n}\rightarrow a_0$. Taking this value for all $n$, we can evaluate \reff{expectation} to find that the mean hitting time is given by
\bea
t_{cl}=\expec{(T)}&=&\sum_{n=d}^{\infty}2na_0\prod_{m=d}^{n-1}(1-a_0)       \nonumber\\
&=&2d-2+\frac{2}{a_0}   \nonumber\\
&\approx&\frac{4}{3}3^d
\label{result}
\eea

The quantum analogue of this mean hitting time is given by the time for perfect state transfer between the
antipodal points $A$ and $B$. We proved in the previous Section
that this time is a constant $t_0= \pi / {\sqrt{2}}$. On comparing
this with \reff{result}, we conclude that the graph $G^d$ provides
an example of a graph for which the quantum case leads to an
exponential separation.

\section{Projecting a hypercube on to a spin chain}
\label{Section-Projection}

\begin{figure}
\begin{center}
\includegraphics[width=6.0cm]{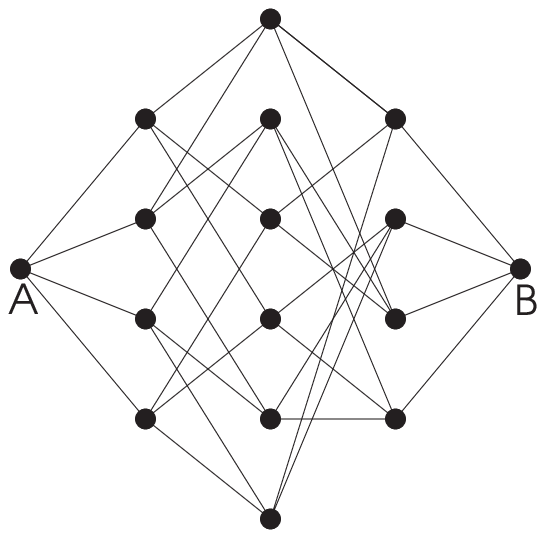}

\caption{An example of a 5--column graph that allows perfect state transfer between either end.}
\label{fig:d=3}
\end{center}
\end{figure}

Encouraged by the ability of the hypercube to allow perfect state
transfer (Section \ref{Section-Greater}), we examine the one--link
hypercube from a different angle. Such a graph falls into a
general category of graphs, $G$, that have the property that the
vertices can be arranged in columns so that there are no edges
between the vertices within any column, and edges only join
vertices in adjacent columns. Further, each vertex in column $i$
must have the same number of incoming (from column $i-1$) and
outgoing (to column $i+1$) edges as all other vertices in that
column. See Figure \reff{fig:d=3} for an example.

Representing the one--link d--dimensional hypercube in such a form, we allow the graph $G$ to consist of $N_C$ columns. The size of each column (the column occupation) is given by $b_i:=
|G_i|=\binom{N_C-1}{i-1}$ and the vertices in each $G_i$ are labelled
$G_{ij}$, $j=\{1, \ldots, b_i\}$. The $i^{th}$ column is $i-1$ edges away from a corner (say $A$) of the hypercube.

The only edges are between vertices of adjacent columns. From each column there must be a set of edges going forwards to the next column, and another set going back to the previous one. These are denoted in the following manner:
\bea
 {\cal P}^{for}_i&\!\!\!:=&\!\!\!\{(G_{ij},k) : j\!\in\!\{1, \ldots, b_i\}, k\!\in\!\{1, \ldots, r_i\} \}   \\
{\cal P}^{back}_i&\!\!\!:=&\!\!\!\{(G_{ij},k) : j\!\in\!\{1, \ldots, b_i\}, k\!\in\!\{1, \ldots, s_i\} \}
\eea
where $r_i$ and $s_i$ denote the number of forward and backward edges respectively for the $i^{th}$ column.
Clearly, if all the edges are to have ends, $|{\cal P}^{for}_i|=|{\cal P}^{back}_{i+1}|$.
Since there is only a single qubit in the first column ($b_1=1$), each vertex in the second column has only a single edge going backwards ($s_2=1$).
With this constraint, and that $s_i$ and $r_i$ must be integers for all $1\leq i\leq N_C$, we require that:
\bea
b_ir_i&=&b_{i+1}s_{i+1} \\
\frac{r_i}{s_{i+1}}&=&\frac{N_C-i}{i}.
\label{cardinality}
\eea
The solution that we will choose for this is $r_i=N_C-i$, $s_i=i-1$, which certainly satisfies all conditions. Thus we have a graph such that for
every pair of numbers $(i,j)$, $G_{ij}$ is connected with $N_C-i$ vertices in
$G_{i+1}$ and each vertex in $G_{i+1}$ is connected with $i-1$ vertices in
$G_{i}$.

Let us define the vectors that span the \emph{column space} $\h{C}$.
\be
\ket{\text{col} \ i} := \frac{1}{\sqrt{b_i}} \sum_{j=1}^{b_i} \ket{G_{ij}}
\ee

Farhi {\em et al.} \cite{ChildsGutmann:2002} note that the evolution with the adjacency matrix $H_G$ of $G$ for this general class of networks (not just the hypercube),
starting in $G_{11}$, always remains in the column space $\h{C}$ because
every vertex in column $i$ is connected to the same number of vertices in
column $i+1$ and every vertex in column $i+1$ is connected to the same number
of vertices in column $i$.

Thus, we can restrict our attention to the column space $\h{C}$ for the
purpose of perfect state transfer from $G_{11}$ to $G_{N_C1}$. The matrix
elements of the adjacency matrix of $G$, restricted to this subspace are
given by
\bea J_i&:=& \bra{\text{col  } i} H_G \ket{\text{col  } i+1} = \sqrt{i (N_C-i)}.    \\
J&=&\left(%
\begin{array}{cccccc}
  0 & J _1 & 0 & 0 & ... & 0 \\
  J_1 & 0 & J_2& 0 &... & 0\\
  0 & J_2 & 0 & J_3 & ... & 0 \\
  0 & 0 & J_3 & 0  & ...& 0 \\
  \vdots &  \vdots &  \vdots & \vdots & \ddots & J_{N_C-1} \\
  0 & 0 & 0 & 0 & J_{N_C-1} & 0\\
\end{array}%
\right).
 \label{hamhopp}
\eea
This can be seen as follows:
\begin{equation}
\begin{split}
  \bra{\text{col } i} H_G \ket{\text{col } i+1}
  \!\!& =\!\!\frac{1}{\sqrt{b_ib_{i+1}}} \!\sum_{j=1}^{b_i} \sum_{j'=1}^{b_{i+1}}
\!\!\bra{G_{i,j}} H_G \ket{G_{i+1,j'}}\\
  & =\!\! \frac{1}{\sqrt{b_ib_{i+1}}} b_i (N_C-i)  =\!\! \sqrt{i(N_C-i)}.\\
\end{split}
\end{equation}

Hence, the above graph exhibits the same behaviour as the $XY$
chain with ``engineered" coupling strengths $J_i$:
\be
H=\half\sum_{i=1}^{N_C-1}J_i(\sigma_i^x\sigma_{i+1}^x+\sigma_i^y\sigma_{i+1}^y).
\ee

 Such a chain must allow perfect state
transfer over any length $N_C$ (where $\ket{A}\equiv\ket{\text{col
} 1}$, $\ket{B}\equiv\ket{\text{col } N_C}$) because the hypercube
does. In the next Section we prove that this is the case using a
more physically motivated derivation.
\par
The number of vertices in the graph, $G$, is given by $|G|=2^{N_C-1}$, hence
it has \emph{communication distance} of $\log_2 |G|$. The two--link hypercube in
contrast has communication distance $2 \log_3 |G|$. One should note however
that the degree of each vertex is bounded linearly.

Some examples of this graph are provided here for different numbers of columns.\\
$N_C=2$: two--qubit chain (d=1 one--link hypercube)\\
$N_C=3$: square (d=2 one--link hypercube)\\
$N_C=5$: for example Figure \reff{fig:d=3} which reduces to an engineered chain, as shown in Figure \reff{fig:spinchain}.

\begin{figure}
\begin{center}
\setlength{\unitlength}{0.85mm}
\begin{picture}(80,27)
\put(20,20){\line(1,0){60}}

\put(20,20){\circle*{2}}

\put(35,20){\circle*{2}}

\put(50,20){\circle*{2}}

\put(65,20){\circle*{2}}

\put(80,20){\circle*{2}}

 \put(0.5,22){\makebox(21,6){$J_n=$}}
 \put(17,22){\makebox(21,6){$\sqrt{1\cdot 4}$}}
 \put(32,22){\makebox(21,6){$\sqrt{2\cdot 3}$}}
 \put(47,22){\makebox(21,6){$\sqrt{3\cdot 2}$}}
 \put(62,22){\makebox(21,6){$\sqrt{4\cdot 1}$}}

 \put(7,12){\makebox(10,5){$n=$}}
 \put(15,12){\makebox(10,5){$1$}}
 \put(30,12){\makebox(10,5){$2$}}
 \put(45,12){\makebox(10,5){$3$}}
 \put(60,12){\makebox(10,5){$4$}}
 \put(75,12){\makebox(10,5){$5$}}

 \put(6.5,6){\makebox(10,5){$m=$}}
 \put(13.5,6){\makebox(10,5){$-2$}}
 \put(28.5,6){\makebox(10,5){$-1$}}
 \put(45,6){\makebox(10,5){$0$}}
 \put(60,6){\makebox(10,5){$1$}}
 \put(75,6){\makebox(10,5){$2$}}
 \put(15,0){\makebox(10,5){$A$}}
 \put(75,0){\makebox(10,5){$B$}}
\end{picture}
\caption{Couplings $J_n$ that admit perfect state transfer from $A$ to $B$ in
a $5$-qubit chain. Eigenvalues $m$ of the equivalent spin-$2$
particle are also shown. This is the projection of Figure \ref{fig:d=3} on to a chain.}\label{fig:spinchain}
\end{center}
\end{figure}

For the purpose of perfect state transfer, we have stated that the $d$--dimensional, one--link hypercube is equivalent to the graph, $G$. The equivalence is obvious for the case of $d=1$ and $d=2$. The general proof arises by considering how the Cartesian product of a graph is taken when you extend the product from $(d-1)$ to $d$ dimensions.

Assume the number of vertices in two adjacent columns are $n_i$ and $n_{i+1}$ in the $(d-1)$--dimensional hypercube. In the $(i+1)^{th}$ column of the d--dimensional hypercube, there must still be the $n_{i+1}$ vertices, plus each of the vertices in the previous column have one more edge (from taking the Cartesian product). Hence the total number of vertices is $n_i+n_{i+1}$. Assuming that the $(d-1)$--dimensional hypercube has column occupations given by a binomial distribution, this specifies that the $d$--dimensional hypercube does as well. Since we know that these column occupations hold for $d=2$, then by induction this must hold for any $d$.

What this does not prove is that the edges between vertices are correct. This is because they aren't necessarily correct. While a hypercube must have a specific set of edges, the construction of the graph $G$ didn't specify which vertices had to be connected to which other ones, we just made sure we got the correct number of forwards and backwards edges. In that sense, the general graph, $G$, is a `scrambled' hypercube. No matter what this scrambling is, $G$ still reduces to the same chain.

\section{State Transfer Over Arbitrary Distances}
\label{Section-Arbitrary}

Suppose we have $N_C$ qubits in a chain, with only one qubit in
state $\ket{\uparrow}\equiv\ket{1}$ and all others in state
$\ket{\downarrow}\equiv\ket{0}$. We previously labelled these as $\ket{j}$, denoting that the single excitation is on the $j^{th}$ qubit. One may associate a fictitious spin $\half
(N_C-1)$ particle with this chain and relabel the basis vectors as
$\ket{m}$, where $m=-\half( N_C-1)+j-1$, as illustrated in Figure \ref{fig:spinchain}. This is an equivalent identification to that made in \cite{Shore}, when considering population transfer between different atomic levels.

The input vertex $\ket{A}$ can be labelled as $\ket{j=1}$ or
$\ket{m=-\half( N_C-1)}$ and the output vertex $\ket{B}$ as
$\ket{j=N_C}$ or $\ket{m=+\half( N_C-1)}$. Now, consider the
Hamiltonian,
\be
H=\lambda
J_x=\half\lambda\sum_{i=1}^{N_C-1}J_i(\sigma_i^x\sigma_{i+1}^x+\sigma_i^y\sigma_{i+1}^y)
\ee
which has the same matrix form as \reff{hamhopp}, with a scaling constant $\lambda$.


This corresponds to the flipped spin hopping between the vertices $j$ and
$j+1$ with a probability amplitude of $J_j$. Now, let us choose $H$
to be proportional to the angular momentum operator $J_x$ or $J_y$
for some spin $J=\half(N_C-1)$ particle. In this case the matrix
elements $J_j$ are $\half\sqrt{j(N_C-j)}$ (these are the same as the elements derived in the previous Section up to a numerical factor). The evolution
of the excitation in the chain is governed by the operator
\begin{equation}
U(t) =\exp \left(-i\lambda t\; J_x\right),
\end{equation}
which represents a rotation of the fictitious spin $J= \half (N_C-1)$
particle. The matrix elements $\bra{j'} U\ket{j}$ are well known.
Thus working out \cite{Feynman} or looking up an appropriate representation of
the $SU(2)$ gives

\begin{equation}
f_{AB}(t)=\bra{B} U(t)\ket{A}=\left[-i\sin\left(\frac{\lambda
t}{2}\right)\right]^{N_C-1}.
\label{eq:engineered_fidelity}
\end{equation}
Thus we get perfect transfer of the state from $\ket{1}$ to
$\ket{N_C}$ in a constant time $t_0=\pi/\lambda$. We can select $N_C-1$ to be divisible by 4 and this eliminates the phase shift caused by the factor of $-i$.

Note that the case of $N_C=2$ is just the same as an unmodulated spin chain of the same length, so the calculation done previously \reff{eq:two_link} is expected to give the same result. This it does, provided we remember that in the current situation the coupling strength is $\lambda/2$, whereas it was simply set to $1$ in the original situation.

Is there any other inter-qubit interaction in the chain that gives
Hamiltonian \reff{hamhopp} when restricted to the single
excitation subspace? The first choice is the $XY$ model with
modulated interactions, another one is the Heisenberg model. If we
try the Heisenberg model of the form

\begin{equation}
\half\sum_{j=1}^{N_C-1} J_j \;\sigma_j\cdot\sigma_{j+1},
\label{heisenberg1}
\end{equation}
we obtain

\begin{equation}
\left(%
\begin{array}{cccccc}
  D_1 & J _1 & 0 & 0 & ... & 0 \\
  J_1 & D_2 & J_2& 0 &... & 0\\
  0 & J_2 & D_3 & J_3 & ... & 0 \\
  0 & 0 & J_3 & D_4  & ...& 0 \\
  \vdots &  \vdots &  \vdots & \vdots & \ddots & J_{N_C-1} \\
  0 & 0 & 0 & 0 & J_{N_C-1} & D_{N_C}\\
\end{array}%
\right)
\end{equation}
where $D_j = \half(\sum_{k=1} J_k) - J_{j-1}-J_j$. In order to get
rid of the diagonal elements in the matrix above we can apply
a magnetic field in the $z$ direction, i.e., we add an extra term to
\reff{heisenberg1},
\begin{equation}
\half\sum_{j=1}^{N_C-1} J_j \;\sigma_j\cdot\sigma_{j+1}
+\sum_{j=1}^{N_C} B_j\sigma^z_j \label{heisenberg2}.
\end{equation}
with $B_j =\half (J_{j-1} + J_j ) - \frac{1}{2(N_C-2)}
\sum_{k=1}^{N_C-1} J_k$.

All this means that we can distribute a quantum state over any
distance with fidelity equal to one as long as we engineer the
inter-qubit interactions, e.g., the inter-qubit distances in the
chain, and apply a suitable spatially varying magnetic field.
\newpage
\section{Scaling Relations and Energy Considerations}

In the previous Section we showed that a spin chain with engineered interactions can be used to transfer a quantum state in fixed time, $t_0$. To compare the computational complexity of the proposed spin chain, it is customary to consider what happens to the energy of the system as the number of spins in the chain increases. One physical assumption that we might make, for example, is that the maximum coupling strength is a fixed size. This maximum occurs at the middle of the chain and is
$$
\lambda J_{\lfloor N/2 \rfloor}\sim\lambda N
$$
Hence, to keep this coupling a constant strength, $\lambda$ must scale with $1/N$ and $t_0=\pi/\lambda$ must scale with $N$.

A second concern is what might happen if we tried to extract our state at a time $t_0-\delta t$. The fidelity of the state transfer is easily approximated from eqn (\ref{eq:engineered_fidelity}) so for small $\delta t$ we get:
$$
f_{AB}(t_0-\delta t)\approx 1-\frac{\pi^2(N_C-1)}{8}\left(\frac{\delta t}{t_0}\right)^2.
$$

Finally, we could ask the question about what happens in the presence of manufacturing errors. In particular, we shall consider what happens if the errors only affect the eigenvalues of the system. This is not the entire story for the spin chain, because we should also consider what happens to the eigenvectors (and, in particular, how well they maintain their symmetry about the centre of the chain since all the eigenvectors are either symmetric or antisymmetric). However, in the case of a double application of the chain (which corresponds to nothing happening to the stored state), we learned in Section \ref{Section-Conditions}, that it is only the eigenvalues that matter.

Let us assume that we have made some manufacturing errors when producing our spin chain i.e., we have some errors that are time independent. The ideal energies of the eigenstates are $E_i$ and the actual energies are $E_i'$.
\begin{eqnarray}
f_{AA}&=&\bra{A}e^{-iH2t_0}\ket{A}	\nonumber\\
\ket{A}&=&\sum_ia_i\ket{i}	\nonumber\\
&=&\sum_ia_ie^{-2it_0E_i}\ket{i}	\nonumber\\
f_{AA}&=&\sum_i|a_i|^2e^{-2it_0(E_i'-E_i)}	\nonumber
\end{eqnarray}
We can estimate the worst case for the fidelity of the identity transformation, by taking the worst error to be $E_i'-E_i=\delta$ and by assuming that $t_0\delta\ll 1$. The error is then
\bea
\epsilon&=&|1-f_{AA}|	\nonumber\\
\epsilon&\approx& 2t_0\delta	\nonumber
\eea
i.e., it scales linearly with $N$.

\section{Using the Chain for Entanglement Transfer}
\label{Section-Entanglement-Transfer}

The idea of the rotation of the large spin particle and subsequent calculation can also tell us more about the system. For example, in the same time that we get perfect state transfer from qubit 1 to $N_C$, we also get perfect state transfer from qubit $j$ to $N_C+1-j$. Under the action of the $J_x$ rotation, these transfers all have the same phase. This means that the chain can be used to move an entangled state from one end of the chain to another. We can start with the Bell state, $1/\sqrt{2}(\ket{01}+\ket{10})$, on the first two qubits:
\begin{equation}
\frac{1}{\sqrt{2}}\left(\ket{1}+\ket{2}\right).
\label{entangle-transfer}
\end{equation}
In time $t_0=\pi/\lambda$ this will evolve to the state
\begin{equation}
\frac{1}{\sqrt{2}}\left(\ket{N_C}+\ket{N_C-1}\right)
\end{equation}
having thus transferred the Bell state to the other end of the
chain. Note that we can not use the state
$1/\sqrt{2}(\ket{00}+\ket{11})$ because this contains a term with
two spins in it, and we have restricted ourselves to the subspace
of only a single spin. We point out, however, that the results of
\cite{Christandl:2004a} show that we will also get perfect state
transfer in higher excitation subspaces and thus, in principle
such a state could be transferred.

The chain can also be used to distribute an entangled pair between two distant parties. If we create a Bell state
\be
\frac{1}{\sqrt{2}}\left(\ket{0}_{NI}\ket{0}_C+\ket{1}_{NI}\ket{1}_C\right)
\label{initial_bell}
\ee
between a non-interacting qubit (NI) and the first qubit on the chain (C), then the overall Hamiltonian will be of the form
\be
H'=\id\otimes H.
\ee
Note that the state $\ket{i}_C$ is exactly the same as the state $\ket{i}$ that were were talking about before with the engineered chain, but we have to be careful not to confuse those states with the states of the non--interacting qubit. The state \reff{initial_bell} then evolves as
\be
\frac{1}{\sqrt{2}}\left(\ket{0}_{NI}e^{-iHt}\ket{0}_C+\ket{1}_{NI}e^{-iHt}\ket{1}_C\right)
\ee
so after the same $t_0$, the entangled pair will be the non-interacting qubit, and the $N_C^{th}$ qubit on the chain.
\be
\frac{1}{\sqrt{2}}\left(\ket{0}_{NI}\ket{0}_C+e^{i\phi}\ket{1}_{NI}\ket{N_C}_C\right)
\ee
This prescription is sufficient to transfer the entanglement of any general two qubit density matrix from being between the non-interacting qubit and the $1^{st}$ qubit at $t=0$ to being between the non-interacting qubit and the $N_C^{th}$ qubit on the chain. This can be understood by seeing how the most general density matrix evolves. What we require is that
\be
\Tr_{\h{G} \setminus \{A\}}\left(\rho(0)\right)=\Tr_{\h{G} \setminus \{B\}}\left(\rho(t_0)\right).
\ee
Such a density matrix can be written as
\bea
\rho(0)&=&\!\!\sum_{(i,j,i',j')\in \{0,1\}}\!\!\alpha_{iji'j'}\ket{ij}\bra{i'j'}    \\
\rho(t)&=&\!\!\sum_{(i,j,i',j')\in \{0,1\}}\!\!\alpha_{iji'j'}e^{-iH't}\ket{ij}\bra{i'j'}e^{iH't}.
\eea
So if a single component of this density matrix evolves, giving perfect transfer, so will all the components and therefore so will the density matrix as a whole. This component evolves as:
\bea
e^{-iH't}\ket{i}_{NI}\ket{j}_C\bra{i'}_{NI}\bra{j'}_Ce^{iH't}   \\
\equiv
\ket{i}_{NI}\left(e^{-iHt}\ket{j}_C\right)\bra{i'}_{NI}\left(\bra{j'}_Ce^{iHt}\right).  \nonumber
\eea
After time $t_0$, if $j$ or $j'$ were 1, then they will have changed to $N_C$, and if they were 0, they remain as 0. Tracing out the effect of all the spins except for the non-interacting one and the $N_C^{th}$ qubit will return precisely the same two qubit density matrix as was initially set up. This then allows the density matrix to be split over the length of the chain.

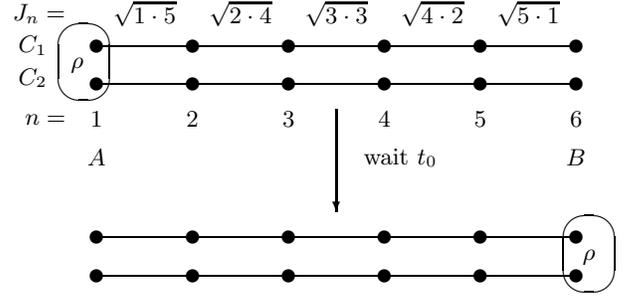
\begin{figure}
\begin{center}
\setlength{\unitlength}{0.85mm}
\begin{picture}(99,38)
\put(17,36){\line(1,0){75}}

\put(17,36){\circle*{2}}
\put(32,36){\circle*{2}}
\put(47,36){\circle*{2}}
\put(62,36){\circle*{2}}
\put(77,36){\circle*{2}}
\put(92,36){\circle*{2}}

 \put(0,38){\makebox(16,6){$J_n=$}}
 \put(14,38){\makebox(21,6){$\sqrt{1\cdot 5}$}}
 \put(29,38){\makebox(21,6){$\sqrt{2\cdot 4}$}}
 \put(44,38){\makebox(21,6){$\sqrt{3\cdot 3}$}}
 \put(59,38){\makebox(21,6){$\sqrt{4\cdot 2}$}}
 \put(74,38){\makebox(21,6){$\sqrt{5\cdot 1}$}}

 \put(2,33){\makebox(10,6){$C_1$}}
 \put(2,28){\makebox(10,6){$C_2$}}

 \put(4,22){\makebox(10,5){$n=$}}
 \put(12,22){\makebox(10,5){$1$}}
 \put(27,22){\makebox(10,5){$2$}}
 \put(42,22){\makebox(10,5){$3$}}
 \put(57,22){\makebox(10,5){$4$}}
 \put(72,22){\makebox(10,5){$5$}}
 \put(87,22){\makebox(10,5){$6$}}

\put(17,30){\line(1,0){75}}
\put(17,30){\circle*{2}}
\put(32,30){\circle*{2}}
\put(47,30){\circle*{2}}
\put(62,30){\circle*{2}}
\put(77,30){\circle*{2}}
\put(92,30){\circle*{2}}

 \put(12,16){\makebox(10,5){$A$}}
 \put(87,16){\makebox(10,5){$B$}}
 \put(9,30.5){\makebox(10,5){$\rho$}}

\put(15,33.5){\oval(8,12)}

\put(17,6){\line(1,0){75}}

\put(17,6){\circle*{2}}
\put(32,6){\circle*{2}}
\put(47,6){\circle*{2}}
\put(62,6){\circle*{2}}
\put(77,6){\circle*{2}}
\put(92,6){\circle*{2}}

\put(17,0){\line(1,0){75}}
\put(17,0){\circle*{2}}
\put(32,0){\circle*{2}}
\put(47,0){\circle*{2}}
\put(62,0){\circle*{2}}
\put(77,0){\circle*{2}}
\put(92,0){\circle*{2}}

 \put(89,0.5){\makebox(10,5){$\rho$}}
\put(94,3.5){\oval(8,12)}

\put(54.5,26){\vector(0,-1){16}}
\put(54.5,16){\makebox(20,5){wait $t_0$}}

\end{picture}
\caption{Scheme for transferring an arbitrary 2--qubit density matrix, $\rho$, using two engineered spin chains ($C_1$ and $C_2$). This example has a chain length of $N_C=6$.}
\label{fig:doublespinchain}
\end{center}
\end{figure}

If we want to transmit the complete density matrix, we just use
two of our engineered chains ($C_1$ and $C_2$) in parallel
($N_{C_1}=N_{C_2}$). The new Hamiltonian can be written as \be
H''=H \otimes H \ee and an exactly analogous argument now applies
so that if we create the desired state (which could be the Bell
state $1/\sqrt{2}(\ket{00}+\ket{11})$, for example) across the
$1^{st}$ qubits of $C_1$ and $C_2$, then after time $t_0$, the
state has been perfectly transmitted to being on the $N_C^{th}$
qubits of the two chains. For an example, see Figure
\ref{fig:doublespinchain}. This scheme will work for both the engineered spin chain and the hypercubes (since the density matrix can be created between the corners of two hypercubes).

\section{$J_y$ and Arbitrary Phase Gates}
\label{Section-Phase}

As previously noted, the $J_x$ rotation introduces a phase shift, depending on the length of the chain. There are several ways in which this can be avoided. The simplest is just to select the correct length of chain. In the case of the engineered chain (and also the one--link hypercube), if $(N_C-1)$ is divisible by 4, then there is no phase shift (since $i^4=1$). Similarly with the two--link hypercube, if the dimension of the hypercube is even, there is no phase shift.

Another choice is to use the $J_y$ rotation (which does not give the factor of $-i$ in \reff{eq:engineered_fidelity}).
\bea
J_y=H=&\half\sum_{j=1}^{N_C-1}J_j(\sigma_j^y\sigma_{j+1}^x-\sigma_j^x\sigma_{j+1}^y)    \\
=&i\left(%
\begin{array}{cccccc}
  0 & -J _1 & 0 & 0 & ... & 0 \\
  J_1 & 0 & -J_2& 0 &... & 0\\
  0 & J_2 & 0 & -J_3 & ... & 0 \\
  0 & 0 & J_3 & 0  & ...& 0 \\
  \vdots &  \vdots &  \vdots & \vdots & \ddots & -J_{N_C-1} \\
  0 & 0 & 0 & 0 & J_{N_C-1} & 0\\
\end{array}%
\right) \nonumber
\eea
Using this in conjunction with the $J_x$ rotation, it is possible, along with the transfer of a state through our spin chain network, to apply an arbitrary phase gate to it during transmission, simply by choosing the correct linear combination of $J_x$ and $J_y$. Assume that we have picked $N_C$ such that $J_x$ gives a phase shift of $i$. A combination of
\begin{equation}
\gamma J_x \pm \sqrt{1-\gamma^2}J_y
\end{equation}
will thus yield a phase shift $e^{i\phi}$ where
\begin{equation}
\tan(\phi)=\frac{\pm\gamma}{\sqrt{1-\gamma^2}}
\end{equation}
meaning that the initial state $\ket{\psi}$ will have evolved to the state
\be
\alpha\ket{0}+e^{i\phi}\beta\ket{N_C}.
\ee

The final alternative for negating the phase shift, or applying an arbitrary phase gate during transmission, would be to apply a uniform global magnetic field in the $z$--direction. Applying a field strength B shifts the energy of the single spin excitation by $B(N_C-2)/2$ and the ground state energy is shifted by $BN_C/2$. Assuming transmission of the state occurs in a time $t_0$, then B can be selected to give the desired phase shift, $\phi$ by
\be
B=\frac{\phi-\frac{\pi}{2}(N_C-1)}{t_0}.
\ee

\section{Summary}
\label{Section-Summary} We have shown that perfect state transfer
is possible across a network of qubits, allowing only control over
the initial design of the network, and no dynamical control.

When the couplings between adjacent qubits are constrained to be
equal, we showed that examples of such networks are the one-- and
two--link $d$--dimensional hypercube. Perfect state transfer for
three- or more-link hypercube geometries is shown to be
impossible. The transfer time is independent of the dimension of
the hypercube and for comparative purposes, we calculated the
expected hitting time in the classical continuous time random
walk, which increases exponentially with the dimension.

We have also proposed a spin chain of $N$ qubits with non-uniform
couplings that allows both state and entanglement transfer. This
chain can be interpreted in two ways: firstly, as a projection of
an $N-1$-dimensional one-link hypercube and secondly, as a
rotation in the $x$-direction of a fictitious spin $(N-1)/2$
particle.

Finally, we have shown how to effect entanglement transfer and how
to introduce phases on the transferred quantum states
\emph{on-the-fly}.

The authors acknowledge financial support from the Cambridge MIT institute. AJL was supported by a Hewlett-Packard Fellowship. MC acknowledges the support of a DAAD Doktorandenstipendium. MC and AK acknowledge the support of the U.K. Engineering and Physical Sciences Research Council.

\end{document}